\documentclass[%
 reprint,
 amsmath,amssymb,
 aps,
 prl,
]{revtex4-1}

\renewcommand\Im{\operatorname{Im}}

\usepackage{color}
\usepackage{graphicx}
\usepackage{bm}

\begin{document}

\title{Fluctuational electrodynamics for nonlinear media}

\author{Heino Soo}
\affiliation{4th Institute for Theoretical Physics, Universit\"at Stuttgart, Germany}
\affiliation{Max Planck Institute for Intelligent Systems, 70569 Stuttgart, Germany}
\author{Matthias Kr\"uger}
\affiliation{4th Institute for Theoretical Physics, Universit\"at Stuttgart, Germany}
\affiliation{Max Planck Institute for Intelligent Systems, 70569 Stuttgart, Germany}
\date{\today}

\begin{abstract}
We develop fluctuational electrodynamics for media with nonlinear optical response. In a perturbative manner, we amend the stochastic Helmholtz equation to describe fluctuations in a nonlinear setting, in agreement with the fluctuation dissipation theorem, and identify the local (Rytov) current fluctuations. We show how the linear response (the solution of the scattering problem) of a collection of objects is found from the individual responses, as measured in isolation. As an example, we compute the Casimir force acting between nonlinear objects which approaches the result for linear optics for large separations, and deviates for small distances.
\end{abstract}

\maketitle

The fluctuating electromagnetic field has become a research area of increasing interest and importance, giving rise to phenomena like Casimir or van der Waals interactions \cite{Casimir48,Milonni} and thermal radiation at far and near fields \cite{Rytov3,Polder71,Eckhardt1984,kubo2012statistical}. Also, the rapid development concerning experimental detection and manipulation, including the framework of MEMS, allows exploration of effects down to the nanoscale \cite{lamoreaux1997demonstration,mohideen1998precision,bressi2002measurement,kittel2005near,rousseau2009radiative,shen2009surface,gad2001mems}. In general, fluctuational electrodynamics has been successfully applied to situations in thermal equilibrium, but also to objects in relative motion or at different temperatures \cite{Pendry97,Polder71,Antezza08}. However, such setups have mostly been considered for media described by linear electric and magnetic responses. 

The field of nonlinear optics is by itself a growing and fundamentally interesting field, comprising, among others, frequency mixing processes, the optical Kerr effect, Brillouin scattering and Raman effects \cite{boyd2003nonlinear}. Especially considering recent developments concerning metamaterials, where large nonlinear response functions are observed, promising novel materials with interesting and useful properties. Examples include media infused with nanoparticles \cite{fukumi1994gold}, organic materials \cite{carter1985time} or polymers \cite{kuebler2000large}.

Fluctuations in nonlinear systems have been investigated for more than 50 years (mostly for classical systems) \cite{van1958thermal,van1965fluctuations}, also in interacting field theories (see e.g. \cite{kardar2007statistical,chang1975quantum}) and applied to critical Casimir forces \cite{krech1994casimir}. Regarding the fluctuating electromagnetic field, short-range enhancement of van der Waals forces have been predicted \cite{makhnovets2016short} and Casimir forces for systems with nonlinear boundary conditions \cite{fosco2015vacuum} and nonlinear coupling functions \cite{kheirandish2011finite} have also been studied. Fluctuations have also been considered in nonlinear optical cavities \cite{drummond1980quantum}, with intriguing effects  regarding heat radiation, studied in the Langevin framework \cite{khandekar2015radiative}. Yet, the direct combination of nonlinear optics and fluctuational electrodynamics is missing in the literature.

In this Letter, we develop fluctuational electrodynamics for systems with nonlinear optical response. Starting from the stochastic nonlinear Helmholtz equation, we develop a perturbative scheme to amend response functions and fluctuations according to the fluctuation dissipation theorem (FDT). We show that the linear response of a system of several objects is not a simple combination of the response functions of the individual objects. Last, we derive and discuss the Casimir force between nonlinear media in equilibrium. We find that an object which is invisible -- in the sense that its linear response is zero -- still feels a Casimir force when brought to a second object due to nonlinear response.


Consider a material described by (linear) dielectric and magnetic responses $\varepsilon$ and $\mu$ as well as a third order nonlinear electric response $\chi^{(3)}$ \footnote{We omit second order nonlinearities for simplicity. These are often absent due to symmetries.}. All response functions can depend on space, thereby allowing the possibility of disconnected objects, e.g. by sharp step functions at the objects' surfaces. The system is -- in Fourier space for time with frequency $\omega$ -- described by a nonlinear Helmholtz equation \cite{boyd2003nonlinear}. In order to include fluctuations, we add a noise source $\mathbf{F}$, whose properties are yet to be determined, 
\begin{equation}
\mathbb{H}\mathbf{E}-\mathcal{N}\left[\mathbf{E}\mathbf{E}\mathbf{E}\right]=\mathbf{F}.\label{eq:SNHE}
\end{equation}
The linear Helmholtz operator is $\mathbb{H}=\nabla\times\nabla\times-\mathbb{V}-\frac{\omega^{2}}{c^{2}}\mathbb{I}$, with the electromagnetic potential $\mathbb{V}=\frac{\omega^{2}}{c^{2}}\varepsilon\left(\omega\right)+\nabla\times\left(\mathbb{I}-\frac{1}{\mathbb{\mu\left(\omega\right)}}\right)\nabla\times$, and speed of light $c$. In what follows, operators (e.g. $\mathbb{H}$ or $\mathbb{V}$) are $3\times3$ matrices and depend on two spatial aguments, such that operator products include matrix multiplication and integration over a joint coordinate. The functional $\mathcal{N}$ describes the third order response, i.e., the $i$th component of $\mathcal{N}\left[\mathbf{E}\mathbf{E}\mathbf{E}\right]$ reads more explicitly,
\begin{eqnarray}
\mathcal{N}\left[\mathbf{E}\mathbf{E}\mathbf{E}\right]_{i} (\omega) =  \frac{\omega^{2}}{c^{2}}\int\mathrm{d}\omega_{1}\mathrm{d}\omega_{2}\mathrm{d}\omega_{3}\,\delta\left(\omega-\omega_{\sigma}\right)\nonumber \\
  \times\chi_{ijkl}^{\left(3\right)}\left(-\omega_{\sigma},\omega_{1},\omega_{2},\omega_{3}\right)\nonumber  E_{j}\left(\omega_{1}\right)E_{k}\left(\omega_{2}\right)E_{l}\left(\omega_{3}\right),
\end{eqnarray}
where $\omega_{\sigma}=\omega_{1}+\omega_{2}+\omega_{3}$ and indices denote spacial components. We assume locality of $\chi^{\left(3\right)}$, so that it couples only fields at equal points in space.
The linear system is solved by the Green's function $\mathbb{G}_{1}$ and the field $\mathbf{E}_{1}$, i.e., $\mathbb{H}\mathbb{G}_{1}=\mathbb{I}$ and $\mathbb{H}\mathbf{E}_{1}=0$. The stochastic equation (\ref{eq:SNHE}) describes the fluctuating electromagnetic field, including quantum- and thermal fluctuations. The noise  $\mathbf{F}$ is chosen such that Eq.~\eqref{eq:SNHE} yields correct expectation values in equilibrium (see Eq.~\eqref{eq:N} below). 
In order to display the FDT in the nonlinear system, we compute response functions and fluctuations, defining $\tilde{\mathbb{G}}$ as the {\it linear response function} of the nonlinear system, i.e., 
\begin{equation}\label{eq:LR}
\tilde{\mathbb{G}}\equiv\left.\frac{\delta\langle \mathbf{E}\rangle }{\delta\mathbf{E}_{\mathrm{in}}}\right|_{\mathrm{eq}}\mathbb{G}_{0},
\end{equation}
with an incoming field $\mathbf{E}_{\mathrm{in}}$ and the vacuum Green's function $\mathbb{G}_{0}$. This linear response function $\tilde{\mathbb{G}}$ obeys FDT \cite{kubo2012statistical}
\begin{equation}\label{eq:FDT2}
\left\langle \mathbf{E}_{\omega}\otimes\mathbf{E}_{\omega^{\prime}}^{*}\right\rangle^{\mathrm{eq}}  =  \delta\left(\omega-\omega^{\prime}\right)b\left(\omega\right)\mathrm{Im}\tilde{\mathbb{G}}.
\end{equation}
Here, $b\left(\omega\right)=\frac{\hbar}{\pi\varepsilon_{0}}\frac{\omega^{2}}{c^{2}}\left[1-e^{-\frac{\hbar\omega}{k_{\mathrm{B}}T}}\right]^{-1}$ gives the strength of the fluctuations, with Planck's constant $\hbar$, temperature $T$, Boltzmann's constant $k_{\mathrm{B}}$, and permittivity of vacuum $\varepsilon_0$. Note that for linear systems, 
$\tilde{\mathbb{G}}=\mathbb{G}_1$, and the familiar FDT \cite{Eckhardt1984} is recovered. Eq.~(\ref{eq:SNHE}) can be formally solved by a so-called Lippmann-Schwinger equation \cite{lippmann1950variational,rahi2009scattering}
\begin{equation}
\mathbf{E}=\mathbf{E}_{1}+\mathbb{G}_{1}\mathbf{F}+\mathbb{G}_{1}\mathcal{N}\left[\mathbf{E}\mathbf{E}\mathbf{E}\right].\label{eq:LS.equation}
\end{equation}
We may treat the linear solution as an incoming field, because, with $\mathbf{E}_{1}=\mathbb{G}_{1}\mathbb{G}_{0}^{-1}\mathbf{E}_{\mathrm{in}}$, the linear response in Eq.~\eqref{eq:LR} may be written as $\tilde{\mathbb{G}}=\left.\frac{\delta\langle \mathbf{E}\rangle }{\delta\mathbf{E}_{1}}\right|_{\mathrm{eq}}\mathbb{G}_{1}$. From Eq.~(\ref{eq:LS.equation}), using, without loss of generality, $\left\langle \mathbf{E}\right\rangle ^{\mathrm{eq}}=0$, and vanishing mean of the noise, $\left\langle \mathbf{F}\right\rangle=0$ \footnote{A finite mean of ${\bf F}$ may be absorbed into the left hand side of Eq.~\eqref{eq:SNHE}.}, we find in first order in $\chi^{(3)}$ for $\tilde{\mathbb{G}}$,
\begin{eqnarray}
\tilde{\mathbb{G}} = \left(\mathbb{I}+\mathbb{G}_{1}\mathbb{N}\right)\mathbb{G}_{1}.\label{eq:Gt}
\end{eqnarray}
The operator $\mathbb{N}$ contains the equilibrium correlation of the field $\mathbf{E}$, which is (perturbatively) expressed using Eq.~\eqref{eq:FDT2}, and reads (again, the spatial $\delta$ function appears because of the locality of the nonlinear response),
\begin{eqnarray}
(\mathbb{N})_{ij}&=& 3\delta^{\left(3\right)}\left(\mathbf{r}-\mathbf{r}^{\prime}\right)\frac{\omega^{2}}{c^{2}}\int_{-\infty}^{\infty}\mathrm{d}\omega^{\prime}b\left(\omega^{\prime}\right)\nonumber \\
 & \times & \chi_{ijkl}^{\left(3\right)}\left({\bf r},-\omega,\omega,\omega^{\prime},-\omega^{\prime}\right)\mathrm{Im}{(\mathbb{G}_1)}_{kl}\left(\mathbf{r},\mathbf{r},\omega^{\prime}\right).\label{eq:N<EE>}
\end{eqnarray}
In Eq.~\eqref{eq:N<EE>}, we have the Green's function of the linear system, $\mathbb{G}_1$, since in first order of $\chi^{(3)}$, the solution of the linear problem multiplies $\chi^{(3)}$.
With Eqs.~\eqref{eq:Gt} and \eqref{eq:FDT2}, the fluctuations of the electric field are fixed and known and may thus readily be used to compute equilibrium quantities such as Casimir forces (see below). We additionally derive the correlation function of the noise ${\bf F}$, which may be relevant for out-of-equilibrium scenarios. From Eq.~\eqref{eq:LS.equation}, it follows to be 
\begin{align}\label{eq:N}
\left\langle \left(\mathbb{G}_{1}\mathbf{F}\right)_{\omega}\otimes\left(\mathbb{G}_{1}\mathbf{F}\right)_{\omega^{\prime}}^{*}\right\rangle ^{\mathrm{eq}} & = \delta\left(\omega-\omega^{\prime}\right)b\left(\omega\right)\mathrm{Im}\mathbb{G}_{1}\\
& + \delta\left(\omega-\omega^{\prime}\right)b\left(\omega\right)\mathbb{G}_{1}\left(\mathrm{Im}\mathbb{N}\right)\mathbb{G}_{1}^{*}\nonumber.
\end{align}
Eq.~\eqref{eq:FDT2} yields fluctuations of the electromagnetic field, and another form of FDT \cite{Rytov3,Eckhardt1984} expresses the fluctuations of local currents in the bodies. Using $\mathrm{Im}\tilde{\mathbb{G}}=-\tilde{\mathbb{G}} \mathrm{Im}[\tilde{\mathbb{G}}^{-1}]\tilde{\mathbb{G}}^*$, and \cite{kruger2012trace} $\Im[\mathbb{V}]=-\Im[\mathbb{G}^{-1}-\mathbb{G}_0^{-1}]$, we may rewrite Eq.~\eqref{eq:FDT2} in lowest order in $\mathbb{N}$
\begin{align}\label{eq:R}
\left\langle \mathbf{E}_{\omega}\otimes\mathbf{E}_{\omega^{\prime}}^{*}\right\rangle^{\mathrm{eq}}  =  \delta\left(\omega-\omega^{\prime}\right)b\left(\omega\right)\tilde{\mathbb{G}}\Im\left[\mathbb{V}+\mathbb{N} -\mathbb{G}_0^{-1}\right]\tilde{\mathbb{G}}^*.
\end{align}
Above, $\Im[\mathbb{V}+\mathbb{N} -\mathbb{G}_0^{-1}]$ is identified as the local Rytov currents \cite{Rytov3}, where $\mathrm{Im}\mathbb{G}_0^{-1}$ is the so-called environment dust \cite{Eckhardt1984,kruger2012trace}, and $\Im[\mathbb{V}+\mathbb{N}]$ are the local Rytov-current fluctuations in the objects. These may now be treated with a local equilibrium approximation to access phenomena out of equilibrium (e.g.~heat transfer). We note that the function $\tilde{\mathbb{G}}$ solves  $\tilde{\mathbb{H}}\tilde{\mathbb{G}}=\mathbb{I}$, with $\tilde{\mathbb{H}}=\nabla\times\nabla\times-\tilde{\mathbb{V}}-\frac{\omega^{2}}{c^{2}}\mathbb{I}$ with the potential\begin{equation}
\tilde{\mathbb{V}}=\mathbb{V}+\mathbb{N}.\label{eq:V.correction}
\end{equation}
The potential $\tilde{\mathbb{V}}$, consistently also appearing in the Rytov-current \eqref{eq:R}, may be seen as the analog of the renormalized mass in interacting field theories \cite{kardar2007statistical}. The second term $\tilde{\mathbb{V}}$ in general depends globally on all points in space through $\mathbb{G}_1$ in Eq.~\eqref{eq:N<EE>}. This leads in general to a nontrivial spatial (and shape) dependence for fluctuation effects.

Eqs.~(\ref{eq:Gt}) and (\ref{eq:N<EE>}) determine the linear response of the nonlinear system, $\tilde{\mathbb{G}}$, to first order in $\chi^{\left(3\right)}$. With the explicit correlator for the noise in Eq.~\eqref{eq:N} and the Rytov currents in Eq.~\eqref{eq:R}, they present fluctuational electrodynamics for the nonlinear system, and hence our first main result.

It is important to note that Eq.~(\ref{eq:Gt}) displays important properties of linear response functions (which may be familiar from the electromagnetic Green's function of linear systems); $\tilde{\mathbb{G}}=\tilde{\mathbb{G}}^{\mathrm{T}}$ due to time reversal symmetry, and $\tilde{\mathbb{G}}^{*}\left(\omega\right)=\tilde{\mathbb{G}}\left(-\omega\right)$ due to realness of the time domain response. Also $\tilde{\mathbb{G}}_{\omega}$ is analytic for $\mathrm{Im}\,\omega>0$, so that Matsubara summation can be used to obtain equilibrium averages. This important manifestation of causality can directly be seen from Eqs.~(\ref{eq:Gt}) and (\ref{eq:N<EE>}) by noting that both $\mathbb{G}_1$ and $\chi^{\left(3\right)}\left(-\omega,\omega,\omega^{\prime},-\omega^{\prime}\right)$ are analytic for $\mathrm{Im}\,\omega>0$ \footnote{Analyticity in $\omega^\prime$ is however not given, since the imaginary part cannot be positive for both $\omega^\prime$ and $-\omega^\prime$.}.

Interestingly, in Eq.~\eqref{eq:N<EE>}, the third order susceptibility appears only in the form $\chi^{\left(3\right)}\left(-\omega,\omega,\omega^{\prime},-\omega^{\prime}\right)$. Therefore, in equilibrium, only a subset of third order processes can contribute, for example the optical Kerr effect or the Raman effect. General frequency mixing processes do not contribute, however.

\begin{figure}
\includegraphics[width=.9\columnwidth]{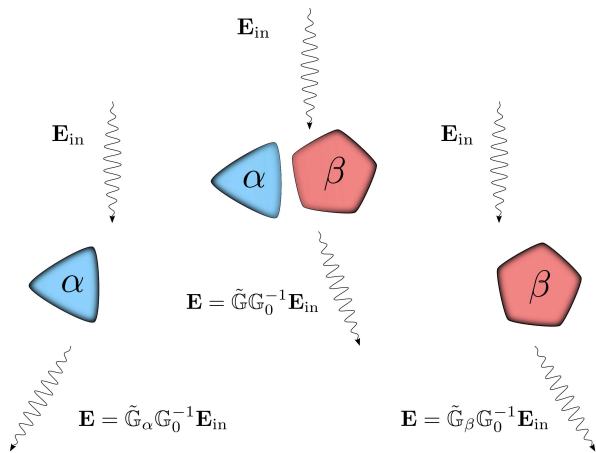}
\caption{Different configurations for a linear response experiment. The electromagnetic potential of nonlinear media is inhomogeneous and dependent on other objects. \label{fig:combination}}

\end{figure}


Turning to a system made of several objects (objects $\alpha$ and $\beta$, see Fig.~\ref{fig:combination}), it is well known that, for purely linear media, the linear response or Green's function of the collection of objects can be found from the results for the isolated objects (see Eq.~\eqref{eq:c} below, which, removing tildes and primes, holds true for linear media). This is the basis of many results found in fluctuational electrodynamics, such as the Lifshitz formula \cite{Lifshitz56}; well known results for Casimir forces or radiative transfer are based on the scattering properties (or $\mathbb{T}$ operators) of the individual objects. In the nonlinear system, this is no longer true: the linear response $\tilde{\mathbb{G}}$ of a collection of objects, as in Fig.~\ref{fig:combination}, takes a nontrivial dependence on the linear responses of the isolated objects ($\tilde{\mathbb{G}}_\alpha$, $\tilde{\mathbb{G}}_\beta$, \dots). Comparing Eq.~\eqref{eq:Gt} for the system containing two objects to its version for isolated objects, we find that 
\begin{equation}
\tilde{\mathbb{G}}=\mathbb{G}^{\prime}+\mathbb{G}^{\prime}\sum_i\left[\mathbb{N}-\mathbb{N}_i\right]\mathbb{G}^{\prime},\label{eq:G.correction}
\end{equation}
where we have introduced the naive (as in the linear case) combination $\mathbb{G}^{\prime}$ for several objects. E.g. , for the case of two objects, we recall (see, e.g., Ref.~\cite{kruger2012trace}),
\begin{align}\label{eq:c}
\mathbb{G}^\prime=\tilde{\mathbb{G}}_{\beta}\frac{1}{\tilde{\mathbb{G}}_{\alpha}+\tilde{\mathbb{G}}_{\beta}-\tilde{\mathbb{G}}_{\alpha}\mathbb{G}_0^{-1}\tilde{\mathbb{G}}_{\beta}}\tilde{\mathbb{G}}_{\alpha}.
\end{align}
The operator $\mathbb{N}$ describes, according to Eq.~\eqref{eq:N<EE>}, the total system, and $\mathbb{N}_i$ describes the situation of object $i$ in isolation.

As mentioned before, Eq.~\eqref{eq:G.correction} states that the linear response of a collection of objects is a nontrivial form of the linear response of the individual ones. This is, again, because fluctuations (zero point and thermal) interact with the incoming field through nonlinearities. This effect is absent for linear media as the incoming and the fluctuating fields are decoupled. It is in principle measurable with scattering experiments, and leads to the different behavior of Casimir forces as described below.


As a concrete application, we compute the Casimir force between two parallel, semi-infinite plates at distance $d$. For linear materials, the well known Lifshitz formula \cite{Lifshitz56} gives the Casimir force for this system in terms of the Fresnel coefficients of the individual plates \cite{dzyaloshinskii1961general}. We give here the result for nonlinear materials. The force, or the Casimir energy, may be found in multiple ways, here we compute the equilibrium correlation function of the electric field in the vacuum between the surfaces, which then gives us the Maxwell stress tensor and the force. 

In the following, we consider homogeneous and isotropic materials, for which the result is given in the appendix. In order to keep the discussion simple, we restrict to frequency independent material parameters. The Lifshitz force then depends on the linear response of the individual plates ($\varepsilon_\alpha$), as well as on the nonlinear function $\chi^{(3)}_\alpha$. Omitting terms of order $(\chi^{(3)})^2$, it suffices to consider only $\chi^{(3)}$ of one of the two plates to be finite (the effect of $\chi^{(3)}$ of the other plate is found by exchanging the plate indices). 

The force evaluation comprises two frequency integrals (see Eq.~\eqref{eq:N<EE>} for the additional integral), both of which are evaluated on the imaginary axis via Matsubara summation. While this is naturally possible for $\omega$ (as mentioned), analyticity for $\Im \omega'>0$ can also be shown if  $\chi^{(3)}$ is frequency independent. The pressure $P$ is then split into two terms, the result of the Lifshitz formula for linear media, and a novel term, resulting from $\chi^{(3)}$. In the zero and infinite temperature limits, these terms can be cast as 
\begin{eqnarray}
P^{T\rightarrow0} & = & \frac{\hbar c}{d^{4}}I_{\mathrm{lin}}^{T\rightarrow0}+\frac{\chi^{\left(3\right)}}{\varepsilon_{0}}\left(\frac{\hbar c}{d^{4}}\right)^{2}I_{\mathrm{nl}}^{T\rightarrow0},\label{eq:P.0}\\
P^{T\rightarrow\infty} & = & \frac{k_{\mathrm{B}}T}{d^{3}}I_{\mathrm{lin}}^{T\rightarrow\infty}+\frac{\chi^{\left(3\right)}}{\varepsilon_{0}}\left(\frac{k_{\mathrm{B}}T}{d^{3}}\right)^{2}I_{\mathrm{nl}}^{T\rightarrow\infty},\label{eq:P.inf}
\end{eqnarray}
where we used that, for isotropic materials, $\chi_{iikk}^{\left(3\right)}	=	\chi_{ikki}^{\left(3\right)}=\chi_{ikik}^{\left(3\right)}=\chi^{\left(3\right)}$ for $i\ne k$ and $\chi_{iiii}^{\left(3\right)}	=	3\chi^{\left(3\right)}$. The functions $I$ are independent of temperature and distance and we note the following  properties of the Casimir force: for $T\to0$, the nonlinear term is proportional to $\hbar^2$, and diverges with $1/d^8$, i.e., it is irrelevant at large distances, and relevant at small $d$. This can be understood intuitively: For small $d$, the intracavity field fluctuations are large, and probe the nonlinear regime. In the high temperature limit, the behavior is very similar, being proportional to $(k_BT)^2$, and diverging with $1/d^6$.

\begin{figure}
\includegraphics[width=1\columnwidth]{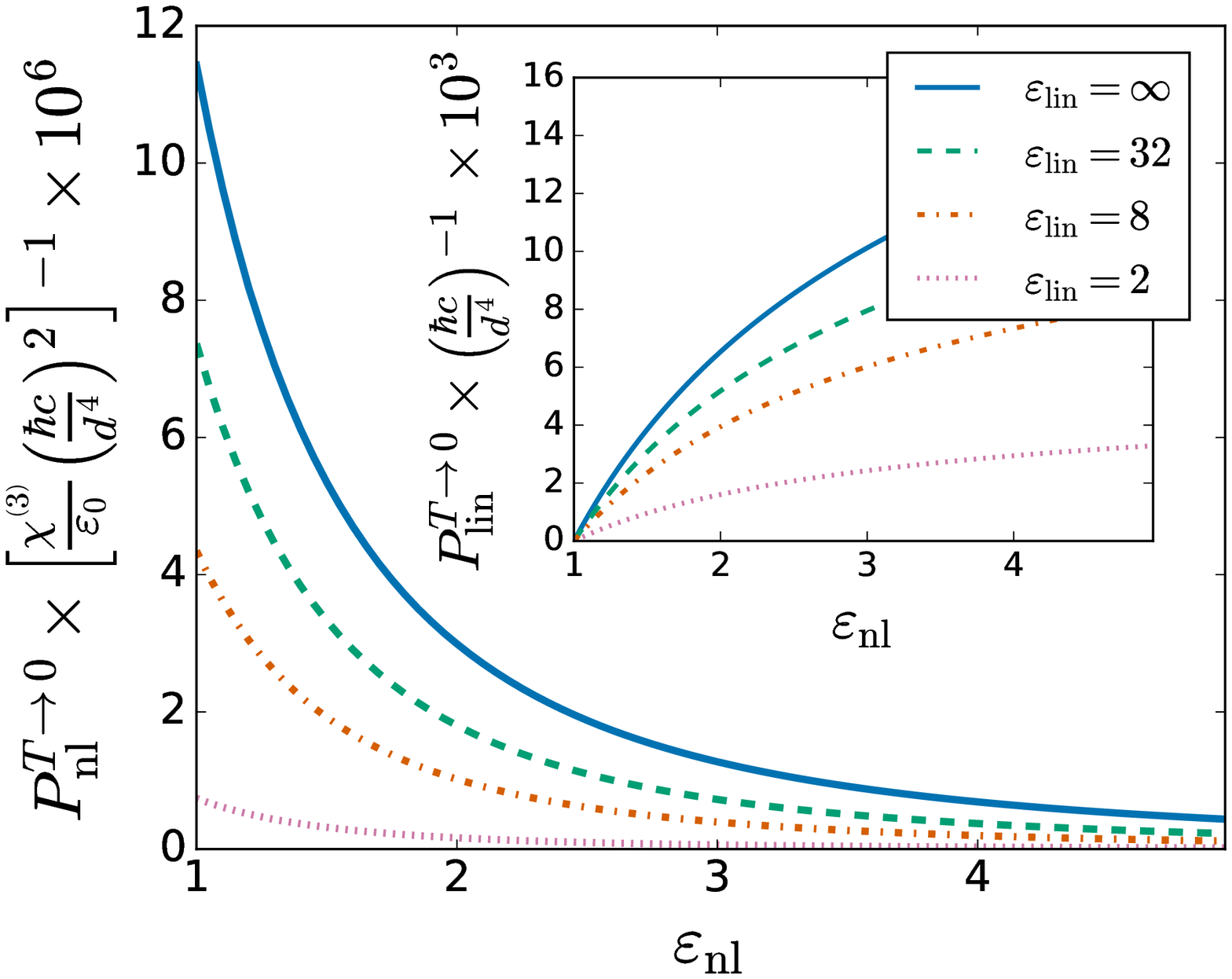}
\caption{Main graph: Nonlinear contribution to the Casimir pressure for different values of $\varepsilon_{lin}$, as labeled, as a function of  $\varepsilon_{nl}$. This graph refers to the quantum limit, $T\to0$. Inset shows the corresponding linear term.\label{fig:Low-temperature-limit}}
\end{figure}

\begin{figure}
\includegraphics[width=1\columnwidth]{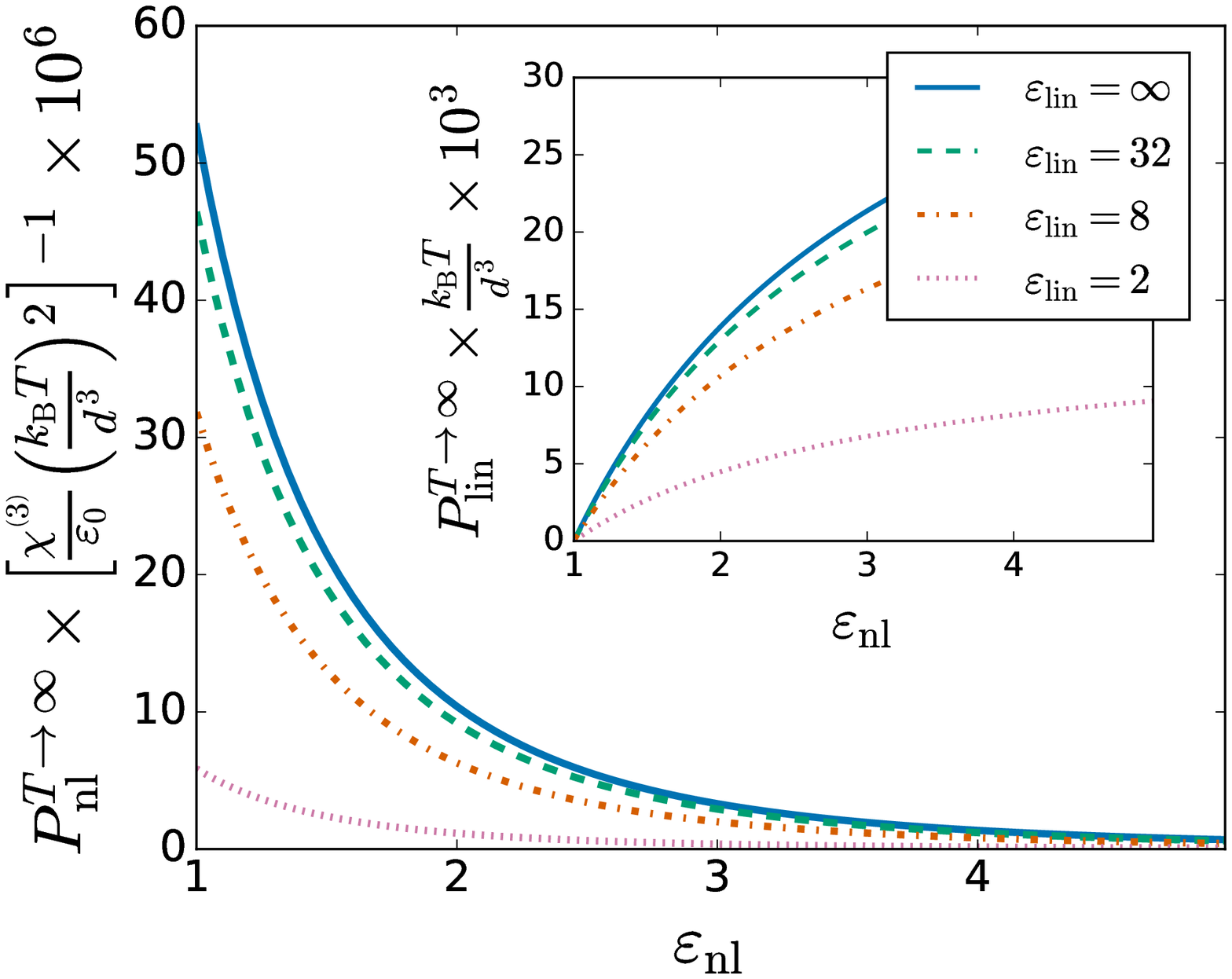}
\caption{Main graph: Nonlinear contribution to the Casimir pressure for different values of $\varepsilon_{lin}$, as labeled, as a function of  $\varepsilon_{nl}$. This graph refers to the thermal limit, $T\to\infty$. Inset shows the corresponding linear term.\label{fig:High-temperature-limit}}
\end{figure}

\begin{figure}
\includegraphics[width=.9\columnwidth]{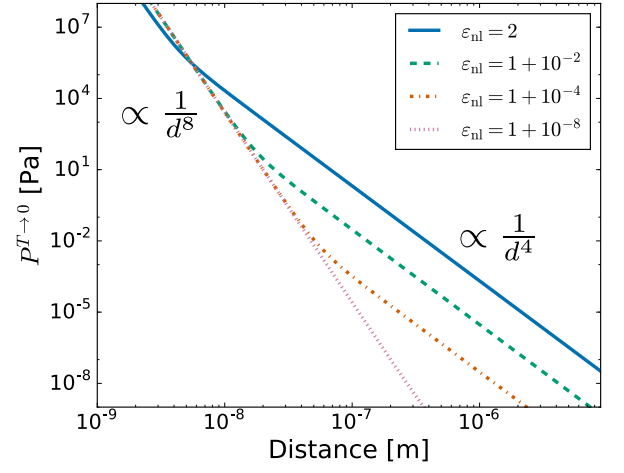}
\caption{Casimir force between a nonlinear object ($\varepsilon_{nl}$ as labeled) and a perfect mirror ($\varepsilon_{\mathrm{lin}}\rightarrow\infty$), in the quantum limit ($T\rightarrow0$), as a function of distance $d$. Here, $\chi^{\left(3\right)}=2\times10^{-16}\,\mathrm{\frac{m^{2}}{V^{2}}}$.\label{fig:Distance-dependence}}
\end{figure}

The numerical results for the functions  $I$ are shown in Figs.~\ref{fig:Low-temperature-limit} and \ref{fig:High-temperature-limit} as functions of the dielectric permittivities $\varepsilon_{\mathrm{lin}}$ (corresponding to the plate with $\chi^{(3)}=0$) and $\varepsilon_{\mathrm{nl}}$ (corresponding to the plate with a finite $\chi^{(3)}$). For a given $\chi^{\left(3\right)}$, the nonlinear pressure contribution is maximal if $\varepsilon_{\mathrm{lin}}\rightarrow\infty$, i.e., if the linear plate is a perfect mirror, and if $\varepsilon_{\mathrm{nl}}\rightarrow1$, i.e., if the the nonlinear plate is transparent. These conditions correspond to minimizing the losses through the linear plate and reflections from the nonlinear plate. As $\varepsilon_{\mathrm{nl}}\rightarrow\infty$, the nonlinear term vanishes. We interpret that in this case, the absorption length in the material vanishes, and the waves cannot penetrate the material to probe nonlinearities.

Figure \ref{fig:Distance-dependence} finally gives the total Casimir force as a function of distance $d$, for the quantum limit ($T\to0$). Here, we have taken $\varepsilon_{lin}\to\infty$. The force takes the well known law of $1/d^4$ for large $d$, and crosses over to $1/d^8$ for small $d$. The nonlinear susceptibility $\chi^{(3)}=2\times10^{-16}\,\mathrm{\frac{m^{2}}{V^{2}}}$ was used, a value measured for glass infused with silver nanoparticles \cite{Karvonen2013}. We note that the crossover takes place at a distance of a few nanometers and is in experimental reach. 

As already apparent from Fig.~\ref{fig:Low-temperature-limit}, the ratio between nonlinear and linear force can be arbitrarily large if $\varepsilon_{nl}$ approaches unity, i.e., if the nonlinear surface becomes transparent. This is also shown in Fig.~\ref{fig:Distance-dependence}, where we depict the total force for different $\varepsilon_{nl}$. As $\varepsilon_{nl}\to 1$, the nonlinear force becomes more and more dominant. 

This has an interesting extreme limit, serving as a simple paradigm for Casimir forces in nonlinear systems. Taking a fully transparent (invisible) object, i.e., having the linear response of vacuum, $\varepsilon_{nl}=1$, and bringing it close to a perfect mirror, the object feels the following total Casimir force,
\begin{align}\label{eq:CT}
&P  =  \frac{3}{2^{8}\pi^{4}}\varepsilon_{0}\mathrm{Re}\iiiint\mathrm{d}\omega\,\mathrm{d}\omega^{\prime}\mathrm{d}q\mathrm{d}q^{\prime}\chi^{\left(3\right)}\left(-\omega,\omega,\omega^{\prime},-\omega^{\prime}\right) 
\nonumber \\ &\times
\frac{a\left(\omega\right)a\left(\omega^{\prime}\right)}{k^{2}k^{\prime2}} N_{q,q^{\prime}}^{\omega,\omega^{\prime}}\left[\frac{e^{2i\left(p+p^{\prime}\right)d}}{\left(p+p^{\prime}\right)p^{\prime}}+\frac{e^{2i\left(p-p^{\prime*}\right)d}}{\left(p-p^{\prime*}\right)p^{\prime}}\right],
\end{align}
where $N_{q,q^{\prime}}^{\omega,\omega^{\prime}}=qq^{\prime}\left[k^{2}\left(4k^{\prime2}-3q^{\prime2}\right)-q^{2}\left(6k^{\prime2}-7q^{\prime2}\right)\right]$ and $a\left(\omega\right)=b\left(\omega\right)-b\left(-\omega\right)$,  with the integral ranges of $[0,\infty]$. We used $k=\omega/c$ and $p=\sqrt{k^{2}-q^{2}}$, analogous for primed variables.

As mentioned, the force in Eq.~\eqref{eq:CT} diverges as $1/d^8$ and $1/d^6$ in the quantum and thermal limits, respectively \footnote{The functional behavior of $\chi^{(3)}$ may lead to deviations from these laws.}. Using metamaterials, such extreme material properties may be approached, e.g., by index matching coating \cite{Southwell1991}. Particles in fluids can also be index matched \cite{budwig1994refractive,wiederseiner2011refractive}.


The combination of fluctuational electrodynamics and nonlinear optics offers a variety of unexplored effects. The stochastic Helmholtz equation is then supplied with an adopted noise strength, allowing description of stochastic processes. In equilibrium, the linear response of a collection of objects is a nontrivial function of the linear responses of the isolated particles, and the Casimir force acting between bodies with nonlinear optical properties is amended at small distances. Future work may investigate path integral formulations of this framework and address nonequilibrium situations.  

We thank M. Kardar, G. Bimonte, D.S. Dean, T. Emig, N. Graham, and R. L. Jaffe for discussions. This work was supported by Deutsche Forschungsgemeinschaft (DFG) grant No. KR 3844/2-1 and MIT-Germany Seed Fund grant No. 2746830.

\pagebreak
\widetext
\section{The full correction to the Lifshitz formula\label{sec:Lifshitz.correction}}

\begin{figure}
\includegraphics[width=.5\columnwidth]{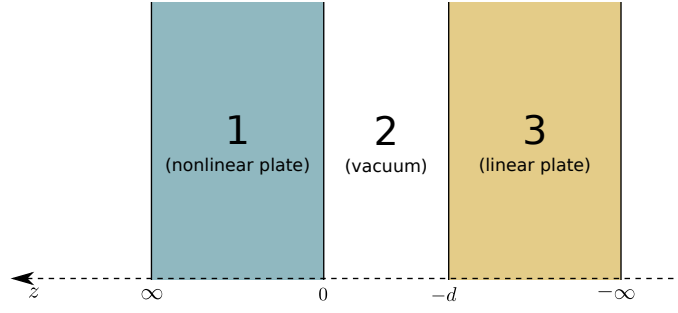}
\caption{Geometry of the system for the Lifshitz formula. \label{fig:geometry}}
\end{figure}

Let us have a linear and nonlinear parallel semi-infinite plates as shown in Figure \ref{fig:geometry}. The correction to the Lifshitz formula is then given by

\begin{eqnarray}
P_{\mathrm{nl}}&=&\frac{3}{2^{8}\pi^{4}}\varepsilon_{0}\mathrm{Re}\iiiint\mathrm{d}\omega\mathrm{d}\omega^{\prime}\mathrm{d}q\mathrm{d}q^{\prime}\,\chi^{\left(3\right)}\left(-\omega,\omega,\omega^{\prime},-\omega^{\prime}\right)\nonumber\\&&\times a\left(\omega\right)a\left(\omega^{\prime}\right)\left[S_{q,q^{\prime}}^{\omega,\omega^{\prime}}\left(d\right)+P_{q,q^{\prime}}^{\omega,\omega^{\prime}}\left(d\right)\right],\\S_{q_{\parallel},q_{\parallel}^{\prime}}^{\omega,\omega^{\prime}}\left(d\right)&=&qq^{\prime}\frac{p_{2}^{2}}{p_{1}^{2}}\mathcal{F}_{23}^{s}\left(\frac{1-\mathcal{F}_{21}^{s}}{1-\mathcal{F}_{21}^{s}\mathcal{F}_{23}^{s}e^{2ip_{2}d}}\right)^{2}\nonumber\\&&\times\left[\frac{e^{2i\left(p_{2}+p_{2}^{\prime}\right)d}}{\left(p_{1}+p_{1}^{\prime}\right)p_{1}^{\prime}}M_{x}\left(\omega^{\prime},q^{\prime},d\right)+\frac{e^{2i\left(p_{2}-p_{2}^{\prime*}\right)d}}{\left(p_{1}-p_{1}^{\prime*}\right)p_{1}^{\prime*}}M_{x}^{*}\left(\omega^{\prime},q^{\prime},d\right)\right],\\P_{q_{\parallel},q_{\parallel}^{\prime}}^{\omega,\omega^{\prime}}\left(d\right)&=&qq^{\prime}\frac{p_{2}^{2}}{p_{1}^{2}}\mathcal{F}_{23}^{p}\left(\frac{1-\mathcal{F}_{21}^{p}}{1-\mathcal{F}_{21}^{p}\mathcal{F}_{23}^{p}e^{2ip_{2}d}}\right)^{2}\nonumber\\&&\times\left[\frac{q_{\parallel}^{2}}{k_{2}^{2}}\left[\frac{e^{2i\left(p_{2}+p_{2}^{\prime}\right)d}}{\left(p_{1}+p_{1}^{\prime}\right)p_{1}^{\prime}}M_{z}\left(\omega^{\prime},q^{\prime},d\right)+\frac{e^{2i\left(p_{2}-p_{2}^{\prime*}\right)d}}{\left(p_{1}-p_{1}^{\prime*}\right)p_{1}^{\prime*}}M_{z}^{*}\left(\omega^{\prime},q^{\prime},d\right)\right]\right.\nonumber\\&&\left.\,-\frac{p_{1}^{2}}{k_{2}^{2}}\left[\frac{e^{2i\left(p_{2}+p_{2}^{\prime}\right)d}}{\left(p_{1}+p_{1}^{\prime}\right)p_{1}^{\prime}}M_{x}\left(\omega^{\prime},q^{\prime},d\right)+\frac{e^{2i\left(p_{2}-p_{2}^{\prime*}\right)d}}{\left(p_{1}-p_{1}^{\prime*}\right)p_{1}^{\prime*}}M_{x}^{*}\left(\omega^{\prime},q^{\prime},d\right)\right]\right],\\M_{x}\left(\omega^{\prime},q^{\prime},d\right)&=&2\left(\frac{\mathcal{F}_{23}^{s\prime}-\mathcal{F}_{21}^{s\prime}\mathcal{F}_{21}^{s\prime}\mathcal{F}_{23}^{s\prime}}{1-\mathcal{F}_{21}^{s\prime}\mathcal{F}_{23}^{s\prime}e^{2ip_{2}^{\prime}d}}\right)+\left(3\frac{q_{\parallel}^{\prime2}}{k_{1}^{\prime2}}-2\right)\left(\frac{\mathcal{F}_{23}^{p\prime}-\mathcal{F}_{21}^{p\prime}\mathcal{F}_{21}^{p\prime}\mathcal{F}_{23}^{p\prime}}{1-\mathcal{F}_{21}^{p\prime}\mathcal{F}_{23}^{p\prime}e^{2ip_{2}^{\prime}d}}\right),\\M_{z}\left(\omega^{\prime},q^{\prime},d\right)&=&\left(\frac{\mathcal{F}_{23}^{s\prime}-\mathcal{F}_{21}^{s\prime}\mathcal{F}_{21}^{s\prime}\mathcal{F}_{23}^{s\prime}}{1-\mathcal{F}_{21}^{s\prime}\mathcal{F}_{23}^{s\prime}e^{2ip_{2}^{\prime}d}}\right)+\left(4\frac{q_{\parallel}^{\prime2}}{k_{1}^{\prime2}}-1\right)\left(\frac{\mathcal{F}_{23}^{p\prime}-\mathcal{F}_{21}^{p\prime}\mathcal{F}_{21}^{p\prime}\mathcal{F}_{23}^{p\prime}}{1-\mathcal{F}_{21}^{p\prime}\mathcal{F}_{23}^{p\prime}e^{2ip_{2}^{\prime}d}}\right),
\end{eqnarray}
where, in the $n$th layer, $p_{n}=\sqrt{k_{n}^{2}-q^{2}}$, with $\mathrm{Im}\left\{ p_{n}\right\} \geq0$, $k_{n}=\varepsilon_{n}\left(\omega\right)\frac{\omega}{c}$, and the Fresnel coefficients are given as $\mathcal{F}_{ln}^{s}=\left(p_{l}-p_{n}\right)/\left(p_{l}+p_{n}\right)$, $\mathcal{F}_{ln}^{p}=\left(\varepsilon_{n}p_{l}-\varepsilon_{l}p_{n}\right)/\left(\varepsilon_{n}p_{l}+\varepsilon_{l}p_{n}\right)$. The quantities $p_{n}^{\prime}$, $k_{n}^{\prime}$, $\mathcal{F}_{ln}^{s\prime}$, and $\mathcal{F}_{ln}^{p\prime}$ are defined in the same way, but using $\omega^{\prime}$ and $q^{\prime}$. The integration ranges are from zero to infinity, and $a\left(\omega\right)=b\left(\omega\right)-b\left(-\omega\right)=\frac{\hbar}{\pi\varepsilon_{0}}\frac{\omega^{2}}{c^{2}}\coth\left(\frac{\hbar\omega}{2k_{\mathrm{B}}T}\right)$.

\vfill

\end{document}